\documentclass[conference,letterpaper]{IEEEtran}
\usepackage{graphicx}
\usepackage{amsmath}
\usepackage{cite}
\usepackage[hidelinks,bookmarks=false,bookmarksopen=false]{hyperref}
\usepackage{xcolor}
\usepackage{float}
\usepackage{amssymb}
\usepackage{makecell}
\usepackage{multirow}
\usepackage{tikz}
\usepackage{booktabs}
\usepackage{pgfplots}
\usetikzlibrary{patterns}
\usepgfplotslibrary{groupplots}
\pgfplotsset{compat=1.16}

\hypersetup{
  pdfpagemode=UseNone,
  pdfstartview=FitH,
  hypertexnames=false
}

\IEEEoverridecommandlockouts

\IEEEoverridecommandlockouts
\IEEEpubid{Accepted to IEEE INFOCOM Workshops 2026 (6G AI-RAN 2026), Tokyo, Japan.\\
This arXiv version is a preprint / author version.}

\begin{document}

\title{SLA-Aware Distributed LLM Inference Across Device-RAN-Cloud}

\author{
    \IEEEauthorblockN{\small {
    Hariz Yet,
    Nguyen Thanh Tam\thanks{The first two authors contributed equally to this work.}, 
    Mao V.~Ngo,
    Lim Yi Shen,
    Lin Wei,
    Jihong Park,
    Binbin Chen,
    and Tony Q.~S.~Quek}
    }
    \IEEEauthorblockA{{\small \{
    {hariz\_yet}, 
    {nguyen\_thanhtam},  
    {vanmao\_ngo}, 
    {yishen\_lim},
    {wei\_lin},
    {jihong\_park}, 
    {binbin\_chen}, 
    {tonyquek}\}
    @sutd.edu.sg}
    }
    \IEEEauthorblockA{
    Singapore University of Technology and Design (SUTD)\\
    }
}

\maketitle

\begin{abstract}
Embodied AI requires sub-second inference near the Radio Access Network (RAN), but deployments span heterogeneous tiers (on-device, RAN-edge, cloud) and must not disrupt real-time baseband processing. We report measurements from a 5G Standalone (SA) AI-RAN testbed using a fixed baseline policy for repeatability. The setup includes an on-device tier, a three-node RAN-edge cluster co-hosting a containerized 5G RAN, and a cloud tier.
We find that on-device execution remains multi-second and fails to meet sub-second budgets. At the RAN edge, SLA feasibility is primarily determined by model variant choice: quantized models concentrate below 0.5\,s, while unquantized and some larger quantized models incur deadline misses due to stalls and queuing.
In the cloud tier, meeting a 0.5\,s deadline is challenging on the measured WAN path (up to 32.9\% of requests complete within 0.5\,s), but all evaluated variants meet a 1.0\,s deadline (100\% within 1.0\,s). Under saturated downlink traffic and up to $N{=}20$ concurrent inference clients, Multi-Instance GPU (MIG) isolation preserves baseband timing-health proxies, supporting safe co-location under fixed partitioning.
\end{abstract}


\section{Introduction}
\label{sec:Intro}

Embodied physical AI requires perception--reason--act loops with strict Service Level Agreement (SLA) latency budgets.
Device-only execution is often power/thermal-limited, while cloud-only execution suffers from variable transport delay.
AI-RAN enables inference placement across device, RAN-edge, and cloud, but co-locating large language model (LLM) / vision-language model (VLM) inference with distributed unit (DU) baseband processing on shared graphics processing units (GPUs) risks violating real-time deadlines without strong isolation.

To study this challenge in a realistic setting, we present an empirical evaluation of Service Level Agreement (SLA)-aware distributed LLM/VLM inference across Device--RAN-edge--Cloud on our 5G Standalone (SA) AI-RAN testbed.
Rather than proposing a new orchestrator, we evaluate a \emph{fixed, conservative baseline policy} and fixed system settings to keep conditions repeatable and to isolate feasibility and tail effects under contention.
Our measurements focus on (i) strict-SLA feasibility across tiers and model variants, and (ii) safety of RAN+AI co-location under hard GPU isolation.
Our contributions are threefold:

\IEEEpubidadjcol 

\begin{itemize}
    \item \textit{AI-RAN testbed evidence for distributed intelligence:}
    We report end-to-end (E2E) measurements for SLA-aware VLM inference across \textbf{Device--RAN/Edge--Cloud} on a 5G SA AI-RAN testbed that co-hosts NVIDIA Aerial 5G low O-DU and vLLM inference on shared GH200 GPUs under {fixed MIG isolation}, with unified RAN's performance metrics, O-Cloud (cloud infrastructure platform) metrics, and E2E metrics.

    \item \textit{Tail-dominated SLA feasibility with minimal observables:}
    Under fixed replay cadence, we quantify \textbf{Hit@$L$} feasibility at strict budgets ($L_P{=}0.5$\,s, $L_M{=}1.0$\,s) and show feasibility is dominated by \textbf{variant choice} and \textbf{tail excursions}. We highlight time-to-first-token (TTFT) as a practical stall/queue proxy that drives deadline misses under tight budgets, suitable for near-real-time control.

    \item \textit{Safe RAN+AI coexistence under contention via hard isolation:}
    Under saturated downlink and up to $N{=}20$ concurrent inference clients, we show fixed MIG partitioning preserves DU timing-health proxies and avoids throughput collapse, supporting deployable co-location in our configuration and motivating hard isolation over conventional GPU multiplexing.
\end{itemize}

By grounding SLA feasibility and coexistence behavior in real hardware measurements, our study provides practical guidance for deploying embodied AI services on AI-RAN infrastructure under strict tail-latency constraints.

\noindent\textbf{Related work.}
AI-RAN systems study co-locating AI workloads at RAN sites while protecting DU real-time deadlines~\cite{AutoRAN,Polese2024AINativeRAN,Polese2025Beyond,AIRANAlliance2024Whitepaper}.
Prior prototypes report concurrent RAN\&AI under partitioning/isolation~\cite{nguyen2025adaptiveaimodelpartitioning,Kundu2025AIRAN,Kholmatov2025AoRA}, while YinYangRAN highlights instability under conventional GPU sharing~\cite{YinYangRAN2024}.
Tiered offloading and efficient edge inference appear in~\cite{xu2025coformercollaboratingheterogeneousedge}, and quantization improves feasibility~\cite{Dettmers2022LlmInt8,Frantar2022GPTQ,LinAWQ}.

\noindent\textbf{Gap and our focus.}
Prior AI-RAN systems emphasize architecture or control, but offer limited repeatable evidence on strict sub-second Hit@$L$ across device/RAN-edge/cloud and on whether hard GPU isolation preserves DU timing-health proxies under concurrent AI load. We address this with a fixed-policy measurement baseline that quantifies tail-driven feasibility and co-location safety under fixed MIG partitioning.

\section{AI-RAN Testbed and SLA-aware Inference Methodology}
\label{sec:setup}

\subsection{AI-RAN Setup}
\label{subsec:testbed}

\begin{figure*}[t]
    \centering
    \includegraphics[width=0.70\linewidth]{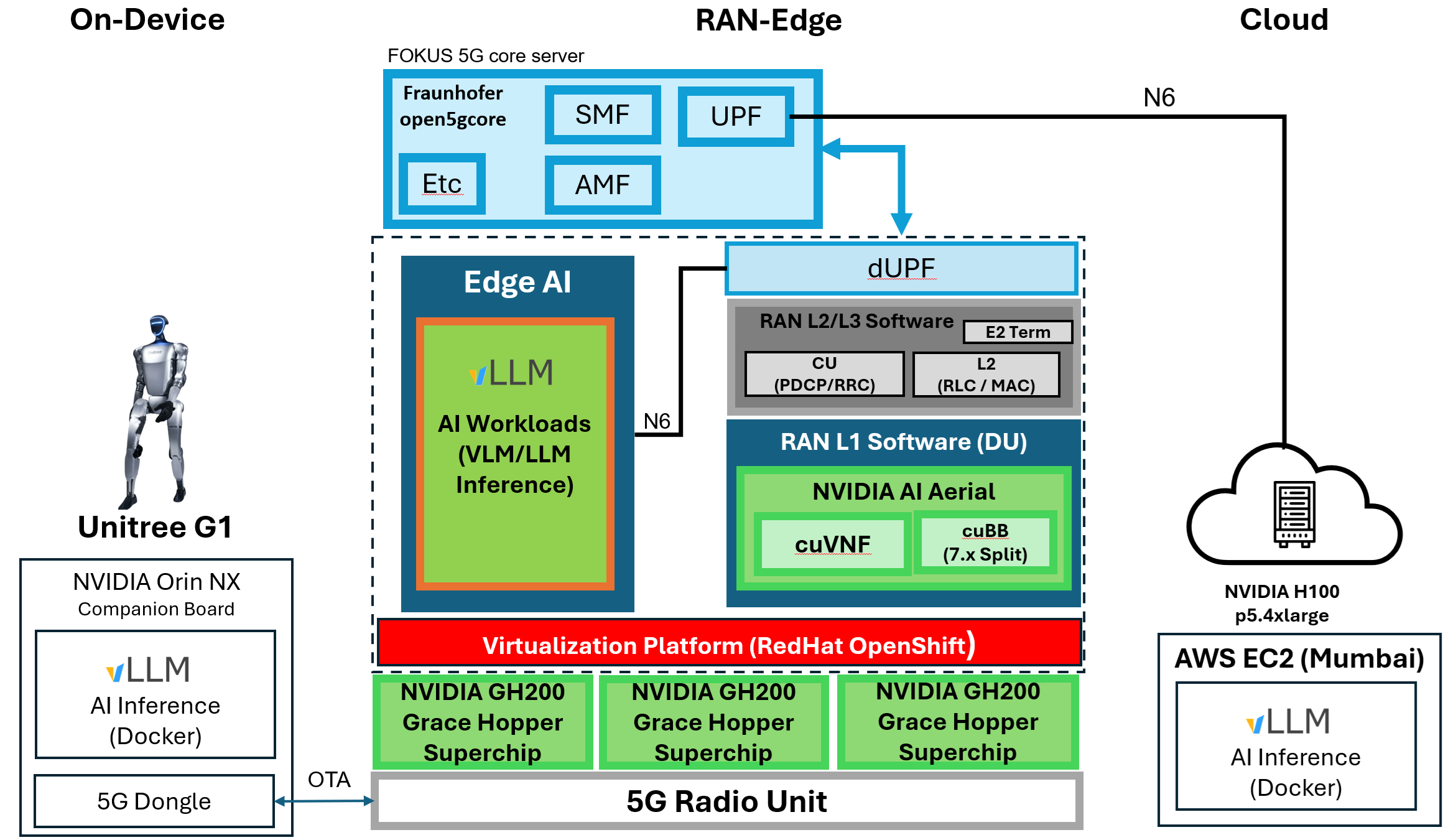}
    \vspace{-6pt}
    \caption{AI-RAN testbed architecture for SLA-tiered distributed LLM inference across Device--Edge--Cloud.}
    \label{fig:architecture}
    \vspace{-10pt}
\end{figure*}

As shown in Fig.~\ref{fig:architecture}, we deploy a 5G SA stack (OpenAirInterface (OAI) O-CU (central unit, CU) / O-DU and NVIDIA Aerial 5G low O-DU) on the GH200 OpenShift cluster, with the radio unit (RU) connected via Split~7.2x fronthaul.
We use Precision Time Protocol (PTP) for RAN timing synchronization and containerized deployment for both RAN and AI components.
GPU resources are managed via the NVIDIA GPU Operator with Multi-Instance GPU (MIG) enabled~\cite{NVIDIAMIG}, pinning Aerial to reserved MIG slices and assigning inference pods to disjoint slices for hardware isolation.
A local breakout distributed User Plane Function (dUPF) enables edge execution without backhaul traversal.

For the device tier, we use a Jetson Orin NX client that can also execute vLLM locally as a Basic-only fallback.
For the cloud tier, we deploy the same serving stack on an AWS Mumbai GPU instance (p5.4xlarge, NVIDIA H100).
\footnote{Cloud results report end-to-end tier experience (serving + transport), i.e., the feasible latency floor and tails observed over the evaluated wide-area network (WAN) path. They are not hardware-normalized server-only comparisons.}
\noindent\textbf{AI-RAN / Open RAN mapping.}
Our stack reflects an AI-native, disaggregated RAN control loop: RAN functions (OAI CU/DU and Aerial low O-DU) export near-real-time telemetry via the O-RAN E2 interface (E2) and Prometheus, while an xApp stores time-synchronized KPIs alongside O-Cloud resource metrics. SLA tiering (Table~\ref{tab:sla}) corresponds to intent-driven service differentiation, and MIG slices provide a hardware-enforced isolation primitive for safe multi-tenant execution at RAN sites.

\subsection{Telemetry and Fixed Baseline Policy}
\textbf{Telemetry and observability:}
We log time-synchronized metrics across (i) the 5G RAN stack, (ii) the O-Cloud platform, and (iii) client-observed service KPIs.
RAN metrics are exported from NVIDIA Aerial (Prometheus) and OAI (E2 agent to FlexRIC near-real-time RAN Intelligent Controller) and stored by an xApp in TimescaleDB.
In the same database we store O-Cloud metrics and inference KPIs, including end-to-end (E2E) latency, round-trip time (RTT), jitter, time-to-first-token (TTFT), and token throughput (TPT), and visualize them via Grafana.

\textbf{Fixed baseline policy (repeatability over optimality):}
We do not implement or evaluate an online orchestrator in this paper.
Instead, we use a fixed, conservative baseline decision flow to keep conditions repeatable and to isolate feasibility and tail effects:
(i) select a model variant based on the SLA budget,
(ii) execute at a chosen tier (on-device/edge/cloud) under availability constraints,
and (iii) pin inference pods to pre-defined MIG slices.
All MIG profiles and placements used for the reported results are fixed throughout each experiment.
The following subsection formalizes the SLA budgets and workload replay used to evaluate feasibility, after which we detail the MIG isolation contract used to protect baseband processing.

\subsection{SLA-aware Distributed LLM Inference}
\label{subsec:sla-model}

This section defines the SLA service model used in our study, and specifies the workload, and stress procedure used to evaluate (i) Strict-SLA feasibility across On Device--RAN--Cloud and (ii) RAN timing health under increasing inference contention. We use a simple, fixed baseline policy to keep experimental conditions repeatable.

\begin{table}[t]
\centering
\footnotesize
\setlength{\tabcolsep}{3.2pt}
\renewcommand{\arraystretch}{1.05}
\begin{tabular}{p{0.18\columnwidth} p{0.18\columnwidth} p{0.56\columnwidth}}
\toprule
\textbf{Tier} & \textbf{Latency Budget($L$)} & \textbf{Guarantees and behavior} \\
\midrule
Premium & $L_P{=}0.5$\,s &
Reserved MIG slice; strict tail-latency target; may preempt lower tiers \\
Medium & $L_M{=}1.0$\,s &
Opportunistic access (quantized variants); served when capacity available \\
Basic & $L_B\geq1.0$\,s &
Best-effort; fallback to on-device/cloud; no exclusivity \\
\bottomrule
\end{tabular}
\vspace{2pt}
\caption{SLA tiers for AI-on-RAN inference services used in this study.}
\vspace{-6pt}
\label{tab:sla}
\end{table}

In our evaluation, we denote the tier E2E latency budget by $L$, with SLA budgets of $L_P{=}0.5$\,s (Premium) and $L_M{=}1.0$\,s (Medium), as detailed in Table~\ref{tab:sla}. Premium services use reserved MIG slices and may pre-empt lower tiers, while Medium and Basic tiers share opportunistic capacity and fall back to other tiers when infeasible.

\noindent\textbf{Why these budgets.}
We use $L_P{=}0.5$\,s and $L_M{=}1.0$\,s to reflect representative perception--action loop timing for embodied navigation: 0.5\,s targets reactive control where late actions reduce safety margin, while 1.0\,s reflects a more tolerant mode where actions remain useful but responsiveness degrades. These budgets also expose the key regime where tail events (queuing stalls and transport variability) dominate feasibility, which is central to AI-RAN placement decisions.

\subsection{MIG Isolation Contract and Enforcement}
\label{subsec:mig-contract}
Our system co-locates AI inference with 5G baseband tasks using NVIDIA MIG, which partitions a single GPU into multiple hardware-isolated slices~\cite{NVIDIAMIG}. MIG enables concurrent execution of Aerial low O-DU processing and vLLM inference on shared GH200 hardware while bounding interference.

\noindent\textbf{Fixed partitioning and placement:}
As shown in Table~\ref{tab:exp_config}, each GH200 GPU (96GB) is partitioned into multiple hardware-isolated MIG instances.
Across the three servers, two nodes use the split $2\times$1g.12GB + $1\times$2g.24GB + $1\times$3g.48GB, while the third node uses $2\times$3g.48GB, with one 3g.48GB slice reserved for Aerial low O-DU.
No MIG instance is shared between Aerial and inference, and MIG profiles remain fixed (no reconfiguration) throughout all runs for all experiments.
Kubernetes device-plugin allocation ensures each pod receives an exclusive MIG slice, enforcing hardware-level isolation between RAN baseband and AI inference workloads.

\noindent\textbf{Tier enforcement:}
To enforce tier priorities, we use Kubernetes \texttt{PriorityClass}: \textbf{Premium} requests are pinned
to reserved MIG slices, while \textbf{Medium/Basic} requests share opportunistic slices and may be preempted when Premium arrives or when RAN guardrails are threatened.

\section{Experimental Setup}
\label{sec:control-loop-architecture}

\subsection{Benchmark task and inference pipeline}
\textbf{Task:}
We emulate a perception--action loop for a humanoid robot (Unitree G1) where the agent outputs discrete navigation commands from egocentric (onboard, first-person) camera frames as input.
Given the latest camera frame(s), the VLM returns one of a small action set
$\mathcal{A}=\{\texttt{FORWARD},\texttt{LEFT},\texttt{RIGHT},\texttt{STOP}\}$ (optionally with a short rationale).
This abstraction captures the core requirement of embodied navigation---timely action selection---without coupling to a specific controller stack.

\textbf{VLM prompt and I/O:}
Each request contains (i) an image extracted from the current video timestamp and (ii) a fixed system prompt that constrains the output format to a single action token (plus optional explanation).
We keep decoding settings constant across tiers and variants (fixed max tokens and temperature) to ensure that latency differences primarily reflect execution tier, quantization, and system load rather than output length variability.
We log the produced action, response size, and token counts for auditability, but our primary objective is \emph{latency feasibility} under strict SLAs rather than task accuracy.

\textbf{Repeatable workload generation:}
To support controlled, apples-to-apples comparisons, we capture a $2.5$-minute first-person video sequence from the robot-mounted camera and replay it with a fixed inter-request interval of $0.5$,s. Each run submits $\approx 300$ requests (one image per request), and we repeat every configuration three times using identical video content and request timing. Edge experiments run on a MIG-partitioned GH200 co-hosting the NVIDIA Aerial low O-DU, while cloud experiments run on AWS Mumbai and include WAN transport. We selected Mumbai because the AWS Singapore region did not offer H100-equipped instances at the time of our experiments, and Mumbai was the nearest region with suitable capacity. More broadly, access to high-end GPU instances varies across regions and countries, which can materially constrain cloud placement decisions.

\subsection{Execution tiers and hardware/software stack}
The three execution tiers and their placement are described in Sec.~\ref{subsec:testbed}. We keep the serving stack and decoding settings consistent across tiers when supported; cloud results are end-to-end (serving plus transport), not server-only latency.

\subsection{Model variants}
We evaluate Qwen2.5-VL (3B/7B) and common formats: unquantized (FP16); activation-aware weight quantization (AWQ); W4A16 (4-bit weights/16-bit activations); and W8A8 (8-bit/8-bit)~\cite{Qwen25VL3BHF}.
We use names such as \texttt{7B-FP16} and \texttt{3B-AWQ}.
Server-side parameters (batching, concurrency limits, and decoding caps) are fixed unless explicitly stated, so results primarily reflect tier placement and model format rather than tuning.

\subsection{Experiment design and measured outcomes}
\textbf{What experiments do we run?}
We run a factorial benchmark over (i) execution tier \{On-Device, Edge, Cloud\}, (ii) model variant \{3B/7B $\times$ FP16/AWQ/W4A16/W8A8\}, and (iii) system load conditions.
For each configuration we replay the same video trace three times and report aggregate statistics.

\textbf{What questions does this answer?}
The experiment matrix is designed to answer:
(1) \emph{Tier feasibility:} which tiers can satisfy sub-second SLAs for embodied AI control?
(2) \emph{Model-format trade-offs:} how much do quantization and parameter count shift deadline hit-rate?
(3) \emph{Tail behavior under contention:} how do queueing and transport variability affect tail latency and deadline misses, especially at the RAN edge when co-hosted with O-DU?

\textbf{Latency metrics and SLA hit-rate.}
We define latency metrics and hit-rate formally in Sec.~\ref{subsec:method} (Latency metrics), and use Hit@$L$ to evaluate feasibility under fixed budgets.

\subsection{Workload, Metrics, and Stress Procedure}
\label{subsec:method}

\begin{table}[t]
\centering
\footnotesize
\setlength{\tabcolsep}{3pt}
\begin{tabular}{p{0.30\columnwidth} p{0.62\columnwidth}}
\toprule
\textbf{Item} & \textbf{Setting (this study)} \\
\midrule
UE device & Jetson Orin NX (client); on-device tier runs execute vLLM locally on Jetson (MAXN) \\
Edge tier & 3$\times$ GH200 (Red Hat OpenShift), NVIDIA H100 MIG enabled; two nodes: 2$\times$1g.12GB + 1$\times$2g.24GB + 1$\times$3g.48GB; one node: 2$\times$3g.48GB with 1$\times$3g.48GB reserved for Aerial \\
RAN stack & OAI O-CU/O-DU + NVIDIA Aerial 5G low O-DU \\
Cloud tier & AWS Mumbai EC2 (p5.4xlarge, NVIDIA H100) running same serving stack \\
Request cadence & 0.5\,s (trace replay), 3 runs per condition \\
Concurrency stress & $N\in\{1,5,10,15,20\}$ parallel clients + saturated downlink (\texttt{iperf}) \\
Metrics logged & E2E latency, TTFT/TPT, hit-rate, RTT, GPU/MIG util, RAN KPIs; {Jetson power rails (CPU/CV + GPU; tegrastats)} \\
Serving stack & vLLM (fixed), CUDA/driver (fixed), Qwen2.5-VL, quant: FP16/AWQ/W4A16/W8A8 \\
MIG profiles & Fixed MIG partitioning; Aerial low O-DU on reserved slices; inference pinned by variant \\
Decoding / tokens & Fixed decoding parameters; token lengths recorded (prompt/output mean/P95) \\
\bottomrule
\end{tabular}
\vspace{2pt}
\caption{Experimental configuration (summary).}
\label{tab:exp_config}
\end{table}

\subsubsection{Latency metrics (tier-agnostic)}
For each request $i$ we measure end-to-end latency $L_i$ as the client-side wall-clock time
from issuing the request until the full response is received.
We also log a serving-stack delay proxy (reported as TTFT) and transport variability when applicable.

In our implementation, the logged \emph{TTFT} is measured as the time from request submission until the first response bytes
are observed at the client. Since responses are streamed, TTFT should be interpreted as a practical indicator of prefill/queue
stalls and service-side contention rather than a true model-internal first-token timestamp.

For \textbf{Edge} and \textbf{Cloud} tiers, we additionally log the per-request round-trip time (RTT) from the underlying TCP
connection (smoothed RTT, SRTT) to characterize transport variability on the evaluated path. For \textbf{Local} execution, RTT
is not applicable and is omitted.

\noindent\textbf{SLA feasibility (hit-rate).}
Given a latency budget $L$ (Sec.~\ref{subsec:sla-model}), we define the deadline hit-rate as
$\mathrm{Hit@}L = \frac{1}{N}\sum_{i=1}^{N}\mathbf{1}[L_i \le L]$.
In our analysis, strict-SLA feasibility is determined by Hit@$L$ and tail behavior; the TTFT proxy serves as a stall/queuing
signal that is strongly associated with deadline misses under tight budgets.

\subsubsection{RAN stress procedure and placements}
To evaluate RAN stability, we run $N \in \{1,5,10,15,20\}$ concurrent inference processes while saturating downlink traffic
using an \texttt{iperf} flow. Each stress condition ($N$ and placement) is executed for one full 2.5-minute trace replay and
repeated three times. We include a no-inference baseline ($N{=}0$) under identical saturated downlink conditions.

We compare two deployments: \textbf{co-located} (vLLM and Aerial on the same node with MIG isolation) and \textbf{different-node}
(vLLM scheduled away from Aerial). We keep RU/DU/CU configuration, UE location, and traffic load constant; only vLLM placement changes.

\section{Experimental Results}
\label{sec:eval}

We aim to answer the following research questions using real hardware measurements:
\textbf{Q1:} How do quantized variants trade off end-to-end latency and prefill delay (TTFT) across On-Device--RAN/Edge--Cloud tiers? 
\textbf{Q2:} How do SLA tiers affect latency and tail behavior under realistic workloads? 
\textbf{Q3:} Can MIG-isolation preserve RAN stability under AI contention?


\begin{table}[t]
\centering
\footnotesize
\setlength{\tabcolsep}{3pt}
\renewcommand{\arraystretch}{1.05}
\begin{tabular}{lcc}
\toprule
\textbf{On-Device} &
\textbf{CPU (W)} & 
\textbf{GPU (W)} \\
\midrule
3B-FP16  & 8.05 & 16.14 \\
3B-AWQ   & 6.00 & 11.29 \\
3B-W4A16 & 6.00 & 11.61 \\
\bottomrule
\end{tabular}
\vspace{2pt}
\caption{On Device-tier power performance on Jetson Orin NX.}
\label{tab:device_local}
\end{table}

\subsection{Latency--Quantization Trade-offs Across Tiers (Q1)}
Quantization yields the largest practical benefit at the \emph{RAN-edge} tier, where it shifts 3B inference from borderline to reliably meeting the \emph{Premium} SLA ($L_P{=}0.5$\,s) while preserving near-100\% feasibility for the \emph{Medium} SLA ($L_M{=}1.0$\,s) (Table~\ref{tab:latency-summary}). In contrast, the \emph{on-device} tier remains multi-second even for 3B variants and therefore misses both strict SLAs (0\% hit-rate at $L_P$ and $L_M$), making it suitable only as a \emph{Basic} fallback. Finally, the \emph{cloud} tier is dominated by end-to-end transport/serving overhead in our setup and does not achieve Premium-grade reliability (Hit@0.5 $\leq 32.9\%$ even on p5.4xlarge), motivating SLA-aware routing toward the RAN-edge whenever available.

\noindent\textbf{On-Device tier is Basic-only under strict SLAs:}
Due to memory constraints, we evaluate only 3B variants on the Jetson Orin NX.
As summarized in Table~\ref{tab:latency-summary}, local execution remains multi-second (E2E $\approx$ 4.7--5.4\,s), yielding 0\% hit-rate at both $L_P{=}0.5$\,s and $L_M{=}1.0$\,s.
Table~\ref{tab:device_local} further shows sustained rail power draw during on-device inference (CPU/CV \textbf{6--8 W}, GPU \textbf{11--16 W})\footnote{\emph{CPU} and \emph{GPU} are Jetson tegrastats domain rail readings (not total input power, and not necessarily additive).},
which is significant for battery-powered robots and competes with onboard perception workloads.
We therefore treat the on-device tier as a \emph{Basic-only} fallback in this study.

\begin{table}[t]
\centering
\footnotesize
\setlength{\tabcolsep}{1.5pt}
\renewcommand{\arraystretch}{1.0}
\begin{tabular}{llcccc}
\toprule
\textbf{\makecell[c]{Variant}} &
\textbf{\makecell[c]{Platform}} &
\textbf{\makecell[c]{Request\ E2E\\Mean$\pm$Std\\(ms)}} &
\textbf{\makecell[c]{TTFT\\Mean$\pm$Std\\(ms)}} &
\textbf{\makecell[c]{RTT\\Mean$\pm$Std\\(ms)}} &
\textbf{\makecell[c]{Hit (\%)\\@0.5/@1.0}} \\
\midrule
\multirow{3}{*}{3B-FP16} & On-Device & 4651$\pm$519 & 353$\pm$447 & -- & 0.0/0.0 \\
                         & Edge      & 490$\pm$35   & 159$\pm$30  & 20.022$\pm$6.138 & 73.9/100.0 \\
                         & Cloud     & 559$\pm$36   & 300$\pm$35  & 84.312$\pm$5.702 & 0.4/100.0 \\
\cmidrule(lr){1-6}
\multirow{3}{*}{3B-AWQ}  & On-Device & 5195$\pm$178 & 352$\pm$15  & -- & 0.0/0.0 \\
                         & Edge      & 391$\pm$29   & 154$\pm$27  & 19.963$\pm$6.235 & 98.3/100.0 \\
                         & Cloud     & 529$\pm$35   & 298$\pm$35  & 83.898$\pm$5.517 & 18.3/100.0 \\
\cmidrule(lr){1-6}
\multirow{3}{*}{3B-W4A16} & On-Device & 5385$\pm$192 & 362$\pm$24 & -- & 0.0/0.0 \\
                          & Edge      & 441$\pm$27   & 157$\pm$24 & 19.751$\pm$6.444 & 97.5/100.0 \\
                          & Cloud     & 562$\pm$35   & 297$\pm$33 & 84.241$\pm$5.643 & 0.3/100.0 \\
\cmidrule(lr){1-6}
\multirow{2}{*}{3B-W8A8}  & Edge  & 428$\pm$31 & 158$\pm$30 & 19.835$\pm$6.527 & 97.1/100.0 \\
                          & Cloud & 520$\pm$30 & 284$\pm$28 & 84.209$\pm$5.555 & 20.3/100.0 \\
\cmidrule(lr){1-6}
\multirow{2}{*}{7B-FP16}  & Edge  & 608$\pm$48 & 162$\pm$26 & 20.910$\pm$6.843 & 0.0/100.0 \\
                          & Cloud & 640$\pm$40 & 323$\pm$30 & 84.278$\pm$5.598 & 0.0/100.0 \\
\cmidrule(lr){1-6}
\multirow{2}{*}{7B-AWQ}   & Edge  & 402$\pm$25 & 154$\pm$23 & 20.249$\pm$6.618 & 99.0/100.0 \\
                          & Cloud & 513$\pm$36 & 314$\pm$36 & 84.132$\pm$5.502 & 32.9/100.0 \\
\cmidrule(lr){1-6}
\multirow{2}{*}{7B-W4A16} & Edge  & 506$\pm$42 & 156$\pm$38 & 19.959$\pm$6.016 & 49.3/99.8 \\
                          & Cloud & 606$\pm$30 & 324$\pm$27 & 84.103$\pm$5.382 & 0.0/100.0 \\
\cmidrule(lr){1-6}
\multirow{2}{*}{7B-W8A8}  & Edge  & 498$\pm$51 & 165$\pm$41 & 20.913$\pm$9.016 & 62.9/99.9 \\
                          & Cloud & 546$\pm$38 & 295$\pm$33 & 83.537$\pm$4.910 & 5.4/100.0 \\
\bottomrule
\end{tabular}
\vspace{2pt}
\caption{E2E, TTFT, RTT, and Hit@0.5/1.0 across tiers for Qwen2.5-VL variants (pooled over 3 runs; $N{=}903$ requests/entry).}
\vspace{-6pt}
\label{tab:latency-summary}
\end{table}

\noindent\textbf{Implications.}
(i) \textit{Premium gating at the edge:} Premium feasibility is largely tail-limited; AWQ remains Premium-feasible while FP16 and some 7B formats exhibit deadline misses under $L_P$.
(ii) \textit{TTFT as a stall signal:} elevated TTFT indicates prefill/queuing stalls that dominate misses under strict budgets.
(iii) \textit{Tier selection:} cloud execution is Medium-feasible but Premium-unreliable on the evaluated WAN path, motivating edge-first routing when available.

\subsection{SLA Feasibility and Hit-Rate Under Tier Budgets (Q2)}
Table~\ref{tab:latency-summary} shows that Premium feasibility at the RAN edge is achieved only by tight-tail variants (notably AWQ and the 3B quantized formats), whereas FP16 and some 7B formats remain tail-limited under $L_P$.
On the evaluated WAN path, cloud execution remains Premium-unreliable but achieves Hit@1.0 $=100\%$ for all variants, indicating that 1.0\,s budgets tolerate the measured transport floor and variability.

\begin{table}[t]
\centering
\footnotesize
\setlength{\tabcolsep}{3.2pt}
\renewcommand{\arraystretch}{1.05}
\begin{tabular}{c r r r r r}
\toprule
$N$ & \multicolumn{3}{c}{\textbf{SlotInd rate (s$^{-1}$)}} & \multicolumn{2}{c}{\textbf{U-plane On-time (\%)}} \\
\cmidrule(lr){2-4}\cmidrule(lr){5-6}
 & Median & P01 & Min & Median & P05 \\
\midrule
0  & 1999.9 & 1998.9 & 1997.9 & 99.984 & 99.970 \\
1  & 2000.0 & 1999.0 & 1997.5 & 99.984 & 99.965 \\
5  & 1999.9 & 1998.9 & 1997.9 & 99.984 & 99.967 \\
10 & 2000.0 & 1999.0 & 1998.7 & 99.984 & 99.964 \\
15 & 1999.9 & 1998.9 & 1997.9 & 99.984 & 99.964 \\
20 & 2000.0 & 1999.0 & 1997.0 & 99.984 & 99.954 \\
\bottomrule
\end{tabular}
\vspace{2pt}
\caption{Timing-oriented baseband health proxies (shared-node, MIG-isolated) under increasing AI contention.}
\label{tab:timing_kpis_aerial}
\vspace{-6pt}
\end{table}

\subsection{RAN Stability Under Co-located and Distributed AI Load (Q3)}
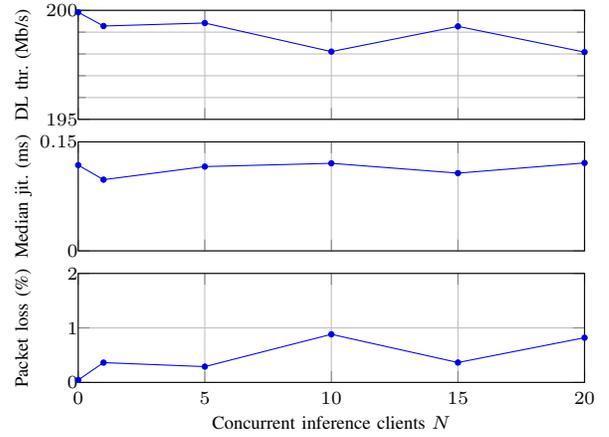
\begin{figure}[t]
\centering
\begin{tikzpicture}
\begin{groupplot}[
    group style={
        group size=1 by 3,
        vertical sep=0.30cm,
        xlabels at=edge bottom,
        xticklabels at=edge bottom,
        ylabels at=edge left,
        yticklabels at=edge left,
    },
    width=0.76\columnwidth,
    height=1.45cm,
    xmin=0, xmax=20,
    xtick={0,5,10,15,20},
    xmajorgrids,
    ymajorgrids,
    grid=both,
    minor x tick num=0,
    tick label style={font=\scriptsize},
    label style={font=\scriptsize},
    scale only axis,
    yticklabel style={text width=2.2em, align=right, inner sep=0pt},
    every axis y label/.append style={
        anchor=center,
        text height=1.7ex,
        text depth=0.3ex,
    },
    every axis x label/.append style={yshift=2pt},
]

\nextgroupplot[
  ylabel={DL thr. (Mb/s)},
  ymin=195, ymax=200,
  ytick={195,200},
  minor y tick num=4,
]
\addplot+[mark=*, mark size=1.0pt] coordinates {
  (0,199.9118) (1,199.28125) (5,199.418226) (10,198.107578) (15,199.265741) (20,198.084617)
};

\nextgroupplot[
  ylabel={Median jit. (ms)},
  ymin=0, ymax=0.15,
  ytick={0,0.15},
]
\addplot+[mark=*, mark size=1.0pt] coordinates {
  (0,0.1180) (1,0.0980) (5,0.1160) (10,0.1205) (15,0.1070) (20,0.1210)
};

\nextgroupplot[
  xlabel={Concurrent inference clients $N$},
  ylabel={Packet loss (\%)},
  ymin=0, ymax=2,
  ytick={0,1,2},
]
\addplot+[mark=*, mark size=1.0pt] coordinates {
  (0,0.044364) (1,0.363096) (5,0.289619) (10,0.882395) (15,0.364787) (20,0.820111)
};

\end{groupplot}
\end{tikzpicture}
\caption{Shared-node RAN KPIs vs. concurrent inference clients $N$ under saturated downlink (MIG-isolated).}

\label{fig:ran_kpis_combined}
\end{figure}

\vspace{-5pt}
We evaluate whether increasing inference contention affects RAN performance.
We first report \emph{radio KPIs} under saturated downlink traffic as $N$ concurrent inference clients increase (Fig.~\ref{fig:ran_kpis_combined}).
We then report \emph{timing-health proxies} that are more directly tied to DU real-time deadlines (Table~\ref{tab:timing_kpis_aerial}).

\noindent\textbf{Radio KPIs (Fig.~\ref{fig:ran_kpis_combined}).}
Across $N \in \{0,1,5,10,15,20\}$, downlink throughput remains close to the 200\,Mb/s target (mean $\approx$198--200\,Mb/s), the \emph{median} jitter stays consistently low (P50 $\approx$0.098--0.121\,ms), and packet loss remains below 1\% (mean $\approx$0.04--0.88\%). No throughput collapse or jitter blow-up is observed over the tested $N$.

\noindent\textbf{Timing-health proxies (Table~\ref{tab:timing_kpis_aerial}).}
SlotInd rate (DU slot-indication events/s) remains tightly concentrated near the expected $2000~\mathrm{s}^{-1}$ for $\mu{=}1$ (0.5\,ms slots), with $\mathrm{P01}\ge 1998.9~\mathrm{s}^{-1}$ for all $N$.
U-plane On-time remains high (P05 $\ge 99.954\%$), indicating stable RU--DU PTP-timed uplink scheduling with no observable degradation under MIG-isolated co-location.

\begin{figure}[t]
\centering
\begin{tikzpicture}
\begin{groupplot}[
    group style={group size=2 by 1, horizontal sep=1.5cm},
    width=0.48\columnwidth,
    height=0.50\columnwidth,
    xtick={0,5,10,15,20},
    grid=both,
    clip=false,
    legend style={
        font=\small,
        draw=none,
        fill=white, fill opacity=0.85, text opacity=1,
        at={(0.5,1.02)}, anchor=south,
        legend columns=2,
    },
]
\nextgroupplot[xlabel={Concurrency $N$}, ylabel={DL BLER$_{95}$ (\%)}, font=\small]
\addplot+[mark=*] coordinates {(0,2.79) (1,5.56) (5,6.69) (10,3.09) (15,6.90) (20,2.92)};
\addlegendentry{Shared-node}
\addplot+[mark=triangle*] coordinates {(0,2.79) (1,1.18) (5,7.01) (10,2.75) (15,8.27) (20,2.74)};
\addlegendentry{Different-node}
\nextgroupplot[xlabel={Concurrency $N$}, ylabel={DL HARQ success (\%)}, ymin=85, ymax=101, font=\small]
\addplot+[mark=*] coordinates {(0,98.40) (1,99.70) (5,91.28) (10,99.46) (15,91.48) (20,99.48)};
\addplot+[mark=triangle*] coordinates {(0,98.40) (1,100.00) (5,92.00) (10,98.78) (15,98.85) (20,98.50)};
\end{groupplot}
\end{tikzpicture}
\caption{MAC-layer tail BLER (left) and HARQ health (right) under increasing AI contention with saturated downlink traffic. $N{=}0$ denotes the no-inference baseline.}
\label{fig:bler_harq_vs_concurrency}
\end{figure}
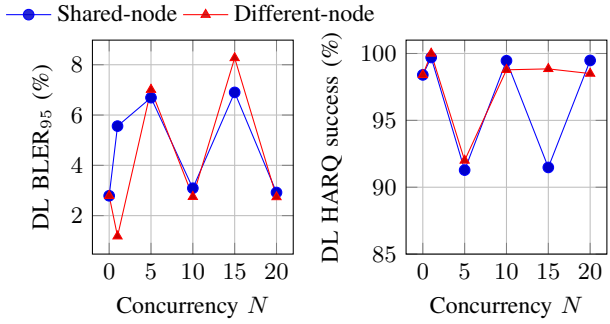

\noindent\textbf{Placement comparison.}
Table~\ref{tab:ran_stability_combined} and Fig.~\ref{fig:bler_harq_vs_concurrency} compare shared-node versus different-node placement; both preserve MAC-layer reliability over the tested $N$. Here, block error rate (BLER) and hybrid automatic repeat request (HARQ) success summarize MAC-layer reliability, while \texttt{iperf} generates a saturated downlink user-plane load. Across the tested range of $N$, both placements preserve radio reliability (BLER/HARQ) without systematic degradation.

\noindent\textbf{Takeaway.}
Under saturated downlink traffic and up to $N{=}20$ concurrent inference clients, fixed MIG partitioning preserves both radio KPIs and timing-health proxies,
supporting safe RAN+AI co-location in our configuration.

\begin{table}[t]
\centering
\footnotesize
\setlength{\tabcolsep}{3.2pt}
\renewcommand{\arraystretch}{1.05}
\begin{tabular}{c r r r r r r}
\toprule
& \multicolumn{3}{c}{\textbf{Shared-node}} & \multicolumn{3}{c}{\textbf{Different-node}} \\
\cmidrule(lr){2-4}\cmidrule(lr){5-7}
$N$ & Mean & BLER$_{95}$ & HARQ & Mean & BLER$_{95}$ & HARQ \\
    & (Mb/s) & (\%) & (\%) & (Mb/s) & (\%) & (\%) \\
\midrule
0  & 199.9 & 2.79 & 98.40 & 199.9 & 2.79 & 98.40 \\
1  & 199.3 & 5.56 & 99.70 & 199.9 & 1.18 & 100.00 \\
5  & 199.4 & 6.69 & 91.28 & 199.8 & 7.01 & 92.00 \\
10 & 198.1 & 3.09 & 99.46 & 199.2 & 2.75 & 98.78 \\
15 & 199.3 & 6.90 & 91.48 & 200.0 & 8.27 & 98.85 \\
20 & 198.1 & 2.92 & 99.48 & 200.0 & 2.74 & 98.50 \\
\bottomrule
\end{tabular}%
\vspace{2pt}
\caption{RAN stability summary under AI contention with saturated downlink.}
\vspace{-6pt}
\label{tab:ran_stability_combined}
\end{table}

\section{Discussion and Limitations}
\label{sec:disc}

\subsection{Why we do not include a ``no-MIG co-location'' baseline}
With MIG disabled, Red Hat OpenShift exposes GPUs as exclusive scalar resources (e.g., \texttt{nvidia.com/gpu});
thus NVIDIA Aerial low O-DU typically claims the full GPU and prevents co-scheduling an inference pod on the same device
(in our cluster, inference pods remain \texttt{Pending} with \texttt{FailedScheduling}/\texttt{Insufficient nvidia.com/gpu})
\cite{NVIDIAGPUOperatorTimeSlicing}. A practical ``no-isolation'' baseline would therefore require enabling GPU oversubscription (CUDA Multi-Process Service (MPS) or CUDA time-slicing), which multiplexes a GPU across pods without hardware memory/fault isolation. From an O-Cloud deployment perspective, the relevant comparison is
\emph{hard isolation} (MIG) versus \emph{soft multiplexing} (MPS/time-slicing). Prior work reports that conventional GPU
sharing can destabilize virtualized RAN (vRAN) timing under contention, motivating hard isolation for production co-location \cite{YinYangRAN2024}.

\subsection{Cloud comparisons and WAN-path dependence}
Cloud results are reported as end-to-end tier experience (serving + transport) rather than as hardware-normalized server-only comparisons.
In our cloud setup (Mumbai p5.4xlarge, NVIDIA H100), the server-side compute gap relative to the GH200 edge tier is reduced; thus remaining Premium misses primarily reflect WAN transport floor and tail excursions on the evaluated path.

In our evaluated setup, cloud execution remains Premium-unreliable across variants (Hit@0.5 $\le 32.9\%$), indicating that the end-to-end remote-execution floor and tail excursions together consume a substantial fraction of the 0.5\,s deadline on the evaluated WAN path.
For Medium budgets, all evaluated variants are feasible in our measurements (Hit@1.0 $=100\%$), suggesting that 1.0\,s budgets can tolerate the observed WAN floor and variability on this route.
Because both the mean transport delay and the tail profile depend on region, routing, and cloud scheduling dynamics, the exact hit-rates are path dependent.
\vspace{-8pt}
\section{Conclusion}
\label{sec:concl}
We measured SLA hit-rate for distributed Qwen2.5-VL inference across device, MIG-isolated RAN-edge, and cloud tiers on a 5G SA AI-RAN testbed co-hosting Aerial low O-DU on GH200.
On-Device tier execution is Basic-only under sub-second SLAs, while at the RAN edge Premium feasibility is achieved only by tight-tail variants (notably AWQ), with FP16 remaining tail-limited.
Cloud execution remains Premium-unreliable on our evaluated WAN path (Hit@0.5 $\le 32.9\%$ even on Mumbai p5.4xlarge), but all evaluated variants are Medium-feasible (Hit@1.0 $=100\%$).
Future work should test additional traces/WAN routes and add admission control that bounds per-slice queueing.

\section*{Acknowledgment}
\label{sec:ack}
This work is supported by the National Research Foundation, Singapore, and the Infocomm Media Development Authority (IMDA) under the Future Communications Research \& Development Programme(FCP). This work was carried out at the Future Communications Connectivity Lab (FCCLab), Singapore University of Technology and Design (SUTD), in collaboration with Red Hat, Inc. The authors thank Michael Tadault, Li Ming Tsai, Danny Yeo, and Shivam Batra (Red Hat, Inc.) for their valuable assistance and contributions. Any opinions, findings, conclusions, or recommendations expressed in this material are those of the authors and do not reflect the views of NRF, Red Hat Inc. and IMDA Singapore.

\bibliographystyle{IEEEtran}
\bibliography{references}

\end{document}